\documentclass[12pt,preprint]{aastex}

\newcommand{\lsim}{\raisebox{-0.3ex}{\mbox{$\stackrel{<}{_\sim} \,$}}}
\newcommand{\gsim}{\raisebox{-0.3ex}{\mbox{$\stackrel{>}{_\sim} \,$}}}
\def\gta{\ifmmode {\mathbin{\lower 3pt\hbox   
    {$\,\rlap{\raise 5pt\hbox{$\char'076$}}\mathchar"7218\,$}}}
    \else {${\mathbin{\lower 3pt\hbox
    {$\rlap{\raise 5pt\hbox{$\char'076$}}\mathchar"7218\,$}}}
    $}\fi}
\def\lta{\ifmmode {\,\mathbin{\lower 3pt\hbox   
    {$\,\rlap{\raise 5pt\hbox{$\char'074$}}\mathchar"7218\,$}}}
    \else {${\mathbin{\lower 3pt\hbox
    {$\rlap{\raise 5pt\hbox{$\char'074$}}\mathchar"7218\,$}}}
    $}\fi}

\shorttitle {Broad Relativistic Iron Line from Ser X-1}
\shortauthors {Bhattacharyya and Strohmayer}

\begin{document}

\title {Evidence of a Broad Relativistic Iron Line from the Neutron
Star Low-Mass X-ray Binary Serpens X-1}

\author {Sudip Bhattacharyya\altaffilmark{1,2}, and Tod
E. Strohmayer\altaffilmark{3}}

\altaffiltext{1}{CRESST and X-ray Astrophysics Lab, Astrophysics
Science Division, NASA's Goddard Space Flight Center, Greenbelt, MD
20771; sudip@milkyway.gsfc.nasa.gov} \altaffiltext{2}{Department of
Astronomy, University of Maryland, College Park, MD 20742}
\altaffiltext{3}{X-ray Astrophysics Lab, Astrophysics Science
Division, NASA's Goddard Space Flight Center, Greenbelt, MD 20771;
stroh@clarence.gsfc.nasa.gov}

\begin{abstract}

We report on an analysis of {\it XMM-Newton} data from the neutron
star low mass X-ray binary (LMXB) Serpens X-1 (Ser X-1). Spectral
analysis of EPIC PN data indicates that the previously known broad
iron K$\alpha$ emission line from this source has a significantly
skewed structure with a moderately extended red wing. The asymmetric
shape of the line is well described with the {\tt laor} and {\tt
diskline} models in XSPEC and strongly supports an inner accretion
disk origin of the line. To our knowledge this is the first strong
evidence of a relativistic line in a neutron star LMXB. This finding
suggests that the broad lines seen in other neutron star LMXBs likely
originate from the inner disk as well. Detailed study of such lines
opens up a new way to probe neutron star parameters and their strong
gravitational fields. The red wing of the iron line from Ser X-1 is
not as broad as that observed from some black hole systems. This is
not unreasonable for a neutron star system, as the accretion disk has
to terminate at or before the hard stellar surface.  Finally, the
inferred source inclination angle in the approximate range $40^{\rm
o}$-$60^{\rm o}$ is consistent with the lack of dips and eclipses from
Ser X-1.

\end{abstract}


\section {Introduction} \label{sec: 1}

The bright persistent LMXB Ser X-1 was discovered in 1965 (Bowyer et
al. 1965).  The detection of type I X-ray bursts from
this source established that it harbors a neutron star (Swank et
al. 1976; Li et al. 1977), as such bursts originate from the
thermonuclear burning of matter on the stellar surfaces (Woosley
\& Taam 1976; Lamb \& Lamb 1978; Strohmayer \& Bildsten 2006).  The
lack of energy dependent dips and eclipses in the X-ray light curve of
Ser X-1 suggests that its inclination angle may be less than $60^{\rm
o}$ (Frank et al 1987; White \& Swank 1982). The analysis of {\it
RXTE} and {\it BeppoSAX} data revealed that the continuum spectra
could be well fitted with a combination of absorbed Comptonization and
disk blackbody models (Oosterbroek et al. 2001). Moreover, these
authors reported the existence of a broad iron emission line near 6
keV, which they could adequately fit with a Gaussian model.  Broad
iron emission lines (near 6 keV) have been observed from many LMXBs
(Asai et al. 2000; Bhattacharyya et al. 2006; Miller et al. 2002a),
and active galactic nuclei (AGN; Reynolds \& Nowak 2003). These lines
could be broadened by either Doppler and relativistic effects (due to
Keplerian motion in the inner accretion disk; Fabian et al. 1989), or
Compton scattering (in a disk corona; Misra \& Kembhavi 1998).  It has
also been suggested recently that complex absorption from perhaps
several ionized components may partially account for some of the broad
line component (e.g., Turner et al. 2005).  The {\it ASCA} data from
the Seyfert-1 galaxy MCG-6-30-15 gave the first strong clues that such
lines originate in the inner accretion disk, as the high
signal-to-noise ratio data revealed a skewed and double-horned
profile, consistent with Doppler and 
relativistic effects in the inner accretion disk, where the speed of
matter is a substantial fraction of the speed of light (Tanaka et
al. 1995). Analysis of more recent data has strongly supported this
inner disk origin of the line for MCG-6-30-15 and other AGN (Reynolds
\& Nowak 2003; Wilms et al. 2001; Reynolds et al. 2004; Reeves et
al. 2007; Miniutti et al. 2007).  Such lines, therefore, are extremely
useful probes of strong gravity, and black hole properties, including
spin (Brenneman \& Reynolds 2006).

An inner disk origin has also been suggested for broad lines in some
stellar mass black hole binaries, such as Cyg X-1 and GRS 1650--500
(Miller et al.  2002a; 2002b; Fabian et al. 1989). But, the origin of
the broad iron lines from neutron star LMXBs has not been as well
understood. This is largely because the available data have a modest
signal-to-noise ratio, which was not precise enough to rule out simple
symmetric profiles (such as a Gaussian profile).  If the inner
accretion disk origin can be established for broad iron lines from
neutron star LMXBs, then the shape and width of the line may be used
to constrain the radius of the inner edge of the disk, as well as the
Keplerian speed at that radius. The former can give an upper limit to
the neutron star radius (as the disk inner edge radius must be greater
than or equal to the stellar radius), while both quantities may be
utilized to constrain the stellar mass.  Moreover, high frequency
quasi-periodic oscillations (kHz QPOs) are observed from many neutron
star LMXBs (van der Klis 2006).  Although the actual physical
mechanism responsible for these timing features is still debated, the
leading models involve strong gravity, with the accretion disk close
to the neutron star surface. Therefore, if the inner disk origin can
be established, the broad iron lines can be useful to constrain kHz
QPO models, and together these two spectral and timing features will
be extremely important as probes of strong gravity, and to constrain
neutron star parameters. We note that the constraints on neutron star
mass and radius can be useful to determine the equation of state (EoS)
of high density cold matter at the stellar core, which is a fundamental
problem of physics (see Bhattacharyya et al. 2005).  In this Letter,
we present evidence for an inner accretion disk origin of the broad
iron line from the neutron star LMXB Ser X-1. In \S~2, we
describe our analysis of {\it XMM-Newton} data, and in \S~3 we discuss
the implications of our results.

\section {Spectral Analysis} \label{sec: 2}

The neutron star LMXB Ser X-1 was observed three times with {\it
XMM-Newton} in March, 2004 (obsIds: 0084020401, 0084020501,
0084020601).  These observations were each separated in time by two
days, and each consists of about 22 ks of data.  Due to the brightness
of the source, there are no European Photon Imaging Camera (EPIC) MOS
detector data available for these observations. EPIC PN data were
obtained in timing mode, and the source photons were also registered
by the Reflection Grating Spectrometer (RGS) instruments. However, as
no significant source spectral lines were detected in the RGS energy
range, here we report only the analysis of the EPIC PN spectra.
We have excluded the portions of the EPIC PN data with high and
variable background (due to soft proton flares), and extracted source
spectra, background spectra, and response matrices using the {\it
XMM-Newton} Science Analysis Software (SAS; version 3.0).  The SAS
task `epatplot' has indicated that this set of spectra (set 1) is
modestly piled up. We have, therefore, extracted another set of
spectra (set 2) after excluding the contributions from the central
(and hence the brightest) pixels in order to minimize the pile-up
effect. Results from the two spectral sets are consistent with each
other, and the broad double-peaked line, which is the focus of this
Letter, is present in both sets in the same energy range. This is
expected, as pile-up cannot generate a double-peaked line. Based on
these findings we do not believe pile-up is a significant concern for
our study.  Therefore, we have used spectral set 1 because of its
higher signal-to-noise ratio.

We have rebinned the spectra (see, for example, Ibarra et al. 2007),
fitted them with various models (within XSPEC), and found that the
best model to describe the observed continuum spectra consists of an
absorbed Comptonization ({\tt compTT}) component plus a disk blackbody
({\tt diskbb}). The addition of {\tt diskbb} to the {\tt compTT}
component is essential, as can be seen from Table 1.  This is in
accordance with observations of Ser X-1 obtained with other X-ray
missions (see Oosterbroek et al. 2001). Oosterbroek et al. noted that
the choice between a disk blackbody and a single temperature blackbody
was arbitrary, however, our spectral fitting shows that a disk
blackbody is preferred to a simple blackbody. For example, the choice
of the former reduces $\chi^2$ by about 47 compared to that for the
latter (obs. 1, i.e., obsId 0084020401). Such a decrease in $\chi^2$
is significant, and a similar decrease is found for obs. 2 (obsId
0084020501) and obs. 3 (obsId 0084020601).  Table 1 shows that the
addition of a Gaussian emission line near 0.54 keV improves the fit
very significantly. As we have not found any source spectral lines at
this energy in the RGS data, we suspect that this feature may be of
instrumental origin, so we have fixed the line centroid of this
feature at $\approx 0.54$ keV in our subsequent model fitting.
Finally, we have found highly significant excess emission near 6
keV. As a broad iron emission line is detected near this energy from
many LMXBs (see \S~1), we have fitted this excess emission with a
Gaussian. This extra component is very significant (see Table 1),
consistent with the results reported in Oosterbroek et
al. (2001). However, as we have mentioned in \S~1, it is not yet known
with any certainty what broadens this line for neutron star LMXBs.  To
explore this question in more detail we have replaced the 6 keV
Gaussian component with the {\tt diskline} model component of XSPEC.
This component represents the spectral line emission from the inner
portion of an accretion disk in the Schwarzschild spacetime (Fabian et
al. 1989).  The corresponding fits show that the {\tt diskline} model
describes the line profile significantly better than the simple
Gaussian profile (see Table 1).  This result supports the notion that
the line is produced in the inner accretion disk.  In order to further
substantiate this result, we have also modeled the spectrum using the
{\tt laor} model. This model is similar to the {\tt diskline} model,
but it includes the effects of the spin of the central star (Laor
1991).  The fact that the {\tt laor} model also fits the line profile
significantly better than the simple Gaussian (see Table 1) provides
strong evidence of the inner disk origin of this line. In fact, {\tt
laor} describes the line profile slightly better than {\tt diskline}
(primarily for obs. 1; see table 1). However, while there is some
indication of a smaller disk inner radius in the {\tt laor} fit using
obs. 1, a value of $6r_{\rm g}$ is not yet excluded with sufficient
significance to argue that the spin of the neutron star is affecting
the line profile.  In principle, deeper observations could test for
such an effect.


In Table 2, we give the best-fit parameter values for the XSPEC model
{\tt wabs*(compTT+ diskbb+gauss+laor)} for each observation.  These
values are consistent across the three observations.  Here we note
that we have used the disk geometry for the Comptonizing plasma ({\tt
compTT} model). However, even the spherical
geometry gives very similar parameter values, except for the {\tt
compTT} optical depth $\tau_{\rm C}$ (the latter geometry gives
$\tau_{\rm C} \approx 15$). We have constrained the rest frame line
energy ($E_{\rm L}$; {\tt laor} component) in the range $6.4-6.97$ keV
while fitting. This is because, K$\alpha$ spectral lines from neutral
or ionized iron are expected in this range.  However, although we have
always found $6.4$ keV as the best-fit value of $E_{\rm L}$, relaxing
this constraint does not lower the best-fit $E_{\rm L}$ much. We have
found the best-fit source inclination angle, $i_{\rm L}$, in the range
$\approx 40^{\rm o}-60^{\rm o}$, which is consistent with the expected
value ($< 60^{\rm o}$) for Ser X-1 (due to the lack of eclipses and
dips; see \S~1). Here we note that, since there are many model
parameters, we have fixed the line parameters (except the
normalizations) to their best-fit values to estimate the errors on the
parameters of the {\tt wabs}, {\tt compTT}, and {\tt diskbb}
components. Similarly, we have frozen the parameters (except the
normalizations) of the {\tt compTT} and {\tt diskbb} models in order
to estimate the errors on the spectral line parameters. Therefore, the
quoted error values are likely somewhat underestimated.
Fig. 1 shows the data, model components, and data-to-model ratio for
observation 2. The {\tt laor} model component for the broad iron line
is clearly shown by the double-peaked dotted line. The moderately
structured shape of the data-to-model ratio results in a relatively
high overall reduced $\chi^2$ for Ser X-1 (Table 2). Note that similar
residuals, and a high reduced $\chi^2$ were also reported by
Oosterbroek et al. (2001; see their Figs. 5 \& 6). These authors
argued that these narrow structures could not be caused by an
incorrect continuum modeling, since such incorrectness would probably
give more smoothly varying data-to-model ratios.  Fig. 2 exhibits the
structure of the broad relativistic iron line. The two panels show
this line from two observations. The data points of the figure clearly
show the line with an extended red wing. This figure also explicitly
shows that both the {\tt laor} model and the {\tt diskline} model
(dotted line in each panel) fit the spectral line profile well.

\section {Discussion and Conclusions} \label{sec: 3}

In this Letter, we report the results of the spectral fitting of the {\it
XMM-Newton} EPIC PN data from the neutron star LMXB Ser X-1. The
best-fit continuum parameter values (see Table 2) are generally
consistent with those of Oosterbroek et al. (2001).  Our slightly
higher Comptonization plasma temperature ($T_{\rm C}$) compared to
that of Oosterbroek et al. can be explained in terms of the lower
$2-10$ keV flux that we have found. This is because, as the intensity
of an LMXB decreases, its energy spectrum generally becomes harder.

It was previously known that Ser X-1 exhibits a broad iron line (see
\S~1), but earlier data did not have the statistical quality to rule
out simple symmetric profiles (such as a Gaussian). As a result, its
origin was not strongly constrained. Knowledge of the line's origin is
very important, because if it is produced in the inner accretion disk,
it can be used (1) to probe the strong gravitational field of the
neutron star, (2) to constrain the stellar parameters, and (3) to
constrain models of the kHz QPOs, which may, in turn, be useful to
achieve the first two goals (see \S~1). By analyzing {\it XMM-Newton}
EPIC PN data we have demonstrated that the line is significantly
asymmetric, and can be accurately modeled with physical models of line
formation in the inner disk (using the {\tt diskline} and {\tt laor}
models in XSPEC; see Fig. 2). This strongly supports the idea that
this spectral line originates from the inner accretion disk of Ser X-1
(see \S~1).  The high signal-to-noise ratio data also revealed the
detailed shape of the broad and relativistically skewed line with an
extended red wing (Fig. 2). Such lines have so far been observed from
AGNs and a few Galactic black hole X-ray binaries (Tanaka et al. 1995;
Miniutti et al. 2007; Miller et al. 2002a; 2002b), but this is the
first strong evidence of a relativistic line in a neutron star
system.  As we have mentioned in \S~1, other neutron star LMXBs
exhibit broad iron emission lines, which so far have been adequately
modeled with simple Gaussian profiles. Our finding for Ser X-1
suggests that the lines seen in other LMXBs may also originate from
the inner accretion disk, and its relativistically skewed nature could
be confirmed with sufficiently long observations.  If future
observations bear this out it will open up an exciting new opportunity
to probe neutron stars and strong gravity with deep spectroscopic
observations.


We have found that the equivalent width ($EW_{\rm L}$) of the iron
line from Ser X-1 is lower than that reported in Oosterbroek et
al. (2001), which may suggest that the strength of this line decreases
with the source intensity. The source inclination angle inferred from
both the {\tt diskline} and {\tt laor} fits is consistent with $\lsim
60^{\rm o}$, which is expected for the non-dipping nature of Ser X-1
(see \S~1).  The best-fit values of $R_{\rm in}$ for obs. 1 \& 3 (see
Table 2) suggest that the accretion disk extends almost to the neutron
star surface. Although this is not the case for obs. 2, it is to be
noted that both the source count rate and the iron line strength are
relatively low for this observation (Table 2), and hence the spectral
line statistics are not as good as for the other two obsIDs. However,
we cannot make a definite conclusion regarding the disk extension with
the current data. But we note that the accretion disk can plausibly
approach the neutron star surface because the stellar magnetic field
in LMXBs is relatively low ($\approx 10^{8 - 9}$ G), and the implied
accretion rate for Ser X-1 is high ($\approx 0.28 -
0.32\dot M_{\rm Edd}$; based on the observed flux, and the Eddington
luminosity $\approx 2.0-3.8\times10^{38}$ ergs s$^{-1}$, for a
distance of $\approx 9.5-12.7$ kpc; Jonker \& Nelemans 2004).  An
important difference between neutron star and black hole systems in
the context of disk line formation is the presence of the neutron star
surface. This sets a firm limit on the inner edge of the disk in a
neutron star system, even a rapidly spinning one, although the exact
limit will depend on the EoS of neutron star matter.  For black holes,
significant spin always implies a smaller disk inner edge radius (for
a corotating disk). Therefore, for very fast spinning black holes, the
red wing of the relativistic iron line is expected to be very
broad. So the relatively modest extended red wing (compared to some
black hole systems; Miniutti et al. 2007) of the iron line from
Ser X-1 appears consistent with the fact that this source harbors a
neutron star. This relativistic iron line is expected to be
accompanied by a continuum reflection spectral component (Reynolds \&
Nowak 2003). A distinct feature of this component is a broad hump near
30 keV. As the signal-to-noise ratio of the EPIC PN data is small at
higher energies ($\gsim 10$ keV), we could not determine if this
component is present in the Ser X-1 spectra.

\acknowledgments

The authors thank Tim Kallman and Jean Cottam for useful discussions,
and John Miller and an anonymous referee for helpful comments.

{}

\clearpage

\begin{deluxetable}{cllll}
\tablecolumns{5}
\tablewidth{0pc}
\tablecaption{Fitting of the {\it XMM-Newton} EPIC PN energy spectra from Ser X-1 
with various XSPEC models.}
\tablehead{No. & XSPEC Model & $\chi^2/{\rm dof}$\tablenotemark{a} & $\chi^2/{\rm
dof}$\tablenotemark{a} & $\chi^2/{\rm dof}$\tablenotemark{a}\\
 & & (Obs. 1\tablenotemark{b}) & (Obs. 2\tablenotemark{b}) & (Obs. 3\tablenotemark{b})}
\startdata
1 & {\tt wabs*compTT} & $\frac{2689.1}{706}$ & $\frac{2517.2}{706}$ & $\frac{2464.3}{706}$ \\ \\
2 & {\tt wabs*(compTT+diskbb)} & $\frac{1708.9}{704}$ (0) & $\frac{1198.0}{704}$ (0) &
$\frac{1432.5}{704}$ (0) \\ \\
3 & {\tt wabs*(compTT+diskbb+gauss)} & $\frac{1517.6}{702}$ (8.0E-19) &
$\frac{1103.8}{702}$ (3.3E-13) & $\frac{1338.2}{702}$ (4.2E-11) \\ \\
4 & {\tt wabs*(compTT+diskbb+gauss+gauss)} & $\frac{1192.7}{699}$ (2.7E-36) &
$\frac{949.1}{699}$ (9.7E-23) & $\frac{1098.0}{699}$ (8.4E-30) \\ \\
5 & {\tt wabs*(compTT+diskbb+gauss+diskline)} & $\frac{1165.6}{696}$ (1.1E-3) &
$\frac{917.4}{696}$ (3.0E-5) & $\frac{1068.9}{696}$ (3.1E-4)
\\ \\
6 & {\tt wabs*(compTT+diskbb+gauss+laor)} & $\frac{1153.6}{696}$ (3.6E-5) &
$\frac{917.8}{696}$ (3.4E-5) & $\frac{1060.5}{696}$ (2.3E-5)
\enddata
\tablenotetext{a}{The number in the parentheses is the probability (calculated from F-test using 
XSPEC) of the decrease of $\chi^2/{\rm dof}$ by chance from the value of the previous row to
that of the current row (for models $2-5$). For model 6, the comparison is with model 4. 
A very small value is replaced with a zero.}
\tablenotetext{b}{Obs. 1: ObsId 0084020401; Obs. 2: ObsId 0084020501; Obs. 3: ObsId 0084020601.}
\end{deluxetable}

\clearpage
\begin{deluxetable}{lccc}
\tablecolumns{4}
\tablewidth{0pc}
\tablecaption{Best fit parameters (with 90\% confidence) for the {\it XMM-Newton} EPIC PN
energy spectra from Ser X-1.}
\tablehead{Parameter\tablenotemark{a} & Obs. 1\tablenotemark{b} & Obs. 2\tablenotemark{b} & Obs.
3\tablenotemark{b}}
\startdata
$N_{\rm H}$\tablenotemark{c} ($10^{22}$ atoms cm$^{-2}$) & $0.48_{-0.02}^{+0.02}$ & 
$0.44_{-0.01}^{+0.01}$ & $0.49_{-0.02}^{+0.02}$ \\
$T_{\rm 0}$\tablenotemark{d} (keV) & $0.17_{-0.01}^{+0.01}$ & $0.19_{-0.01}^{+0.01}$ &
$0.17_{-0.01}^{+0.02}$ \\
$T_{\rm C}$\tablenotemark{e} (keV) & $3.18_{-0.36}^{+0.54}$ & $2.63_{-0.11}^{+0.14}$ &
$2.81_{-0.18}^{+0.35}$ \\
$\tau_{\rm C}$\tablenotemark{f} & $6.19_{-0.74}^{+0.68}$ & $7.25_{-0.32}^{+0.30}$ &
$6.71_{-0.52}^{+0.48}$ \\
$T_{\rm in}$\tablenotemark{g} (keV) & $1.71_{-0.12}^{+0.10}$ & $1.21_{-0.04}^{+0.04}$ &
$1.40_{-0.08}^{+0.08}$ \\
$E_{\rm L}$\tablenotemark{h} (keV) & $6.40_{-0.0}^{+0.08}$ & $6.40_{-0.0}^{+0.03}$ &
$6.40_{-0.0}^{+0.04}$ \\
$\beta_{\rm L}$\tablenotemark{i} & $2.17_{-0.13}^{+0.30}$ & $2.38_{-0.56}^{+0.31}$ &
$2.55_{-0.30}^{+0.39}$ \\
$R_{\rm in}$\tablenotemark{j} & $4.04_{-0.68}^{+2.14}$ & $16.19_{-3.19}^{+8.77}$ &
$5.59_{-1.19}^{+1.48}$ \\
$R_{\rm out}$\tablenotemark{k} & $271.5_{-93.7}^{+115.2}$ & $400.0_{-70.9}^{+0.0}$ &
$313.5_{-166.9}^{+86.5}$ \\
$i_{\rm L}$\tablenotemark{l} (degree) & $44.7_{-3.2}^{+1.7}$ & $50.2_{-5.4}^{+8.8}$ &
$39.7_{-1.5}^{+1.4}$ \\
$EW_{\rm L}$\tablenotemark{m} (eV) & $105.1_{-8.0}^{+6.8}$ & $85.9_{-10.8}^{+8.9}$ &
$95.2_{-8.9}^{+11.6}$ \\ \hline
${\chi^2}/{{\rm dof}}$ & ${1153.6}/{696}$ & ${917.8}/{696}$ & ${1060.5}/{696}$ \\
Flux\tablenotemark{n} ($0.5-2$ keV) & $0.93$ & $0.87$ & $0.91$ \\
Flux\tablenotemark{n} ($2-10$ keV) & $4.22$ & $3.34$ & $3.79$ \\
\enddata
\tablenotetext{a}{Parameters of the XSPEC model {\tt wabs*(compTT+diskbb+gauss+laor)}.}
\tablenotetext{b}{Obs. 1: ObsId 0084020401; Obs. 2: ObsId 0084020501; Obs. 3: ObsId 0084020601.}
\tablenotetext{c}{Hydrogen column density from the {\tt wabs} model component.}
\tablenotetext{d}{Input soft photon (Wien) temperature of the {\tt compTT} model component.}
\tablenotetext{e}{Temperature of the Comptonizing plasma ({\tt compTT} model
component; disk geometry).}
\tablenotetext{f}{Optical depth of the Comptonizing plasma ({\tt compTT} model component).}
\tablenotetext{g}{Temperature at inner disk radius from {\tt diskbb} model component.}
\tablenotetext{h}{Rest frame energy of the broad relativistic iron emission line ({\tt laor} model component).}
\tablenotetext{i}{Power law index of emissivity ({\tt laor} model component).}
\tablenotetext{j}{Inner radius (in the unit of $GM/c^2$; $M$ is the mass of the neutron star) 
from the {\tt laor} model component.}
\tablenotetext{k}{Outer radius (in the unit of $GM/c^2$) from the {\tt laor} model component.}
\tablenotetext{l}{Source inclination angle from the {\tt laor} model component.}
\tablenotetext{m}{Equivalent width of the broad relativistic iron emission line ({\tt laor} model component).}
\tablenotetext{n}{Observed flux in $10^{-9}$ ergs cm$^{-2}$ s$^{-1}$.}
\end{deluxetable}

\clearpage

\begin{figure}
\hspace{-1.9 cm}
\epsscale{0.7}
\plotone{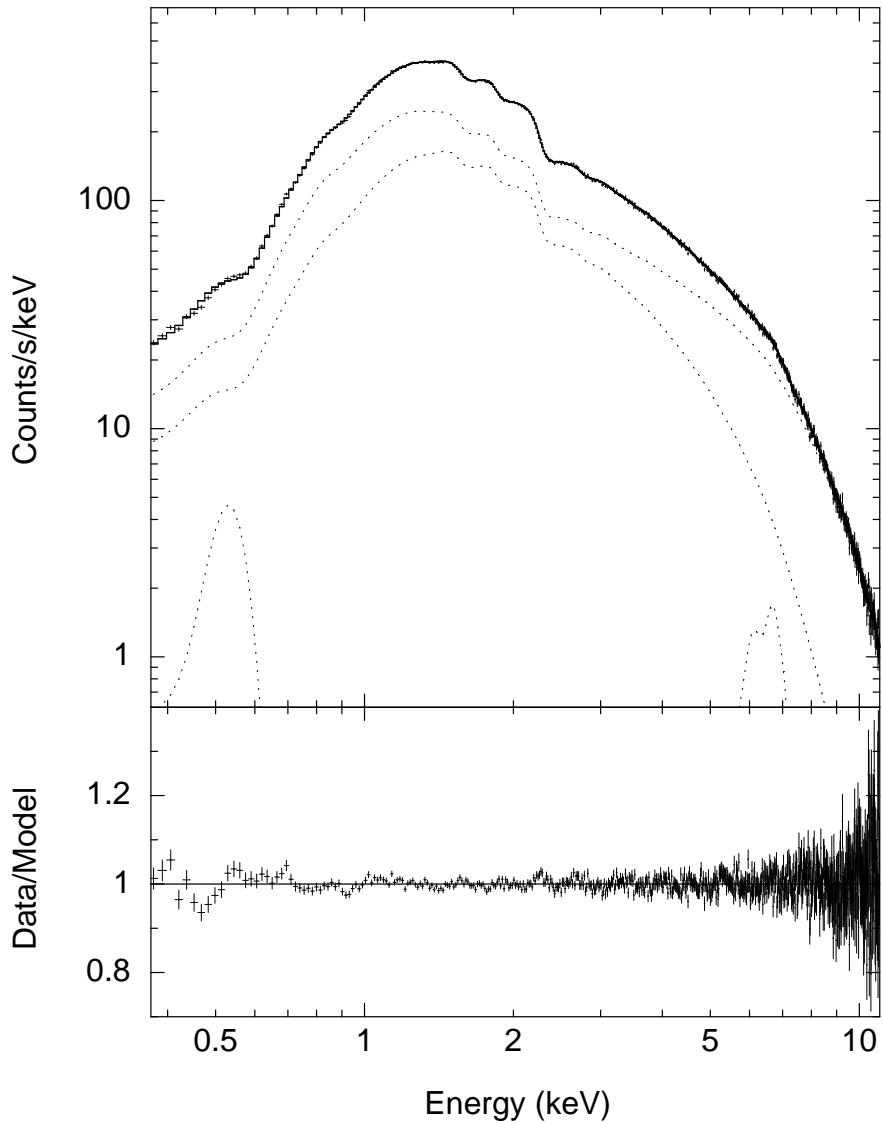}
\vspace{-0 cm}
\caption {{\it XMM-Newton} EPIC PN energy spectrum from Ser X-1 for
obs. 2 (obsId 0084020501).  The upper panel shows the data (error
bars), model (solid line), and individual additive model components
(dotted lines). Here we have used the best-fit XSPEC model {\tt
wabs*(compTT+diskbb+gauss+laor)} (see Table 2). Among the two
continuum additive model components, the upper dotted line shows {\tt
compTT}, and the lower dotted line shows {\tt diskbb}.  The low energy
emission line is the {\tt gauss} component, while the high energy
broad iron emission line is the {\tt laor} component. The lower panel
shows the data to model ratio.}
\end{figure}

\clearpage
\begin{figure}
\hspace{-0.3 cm}
\epsscale{0.4}
\plotone{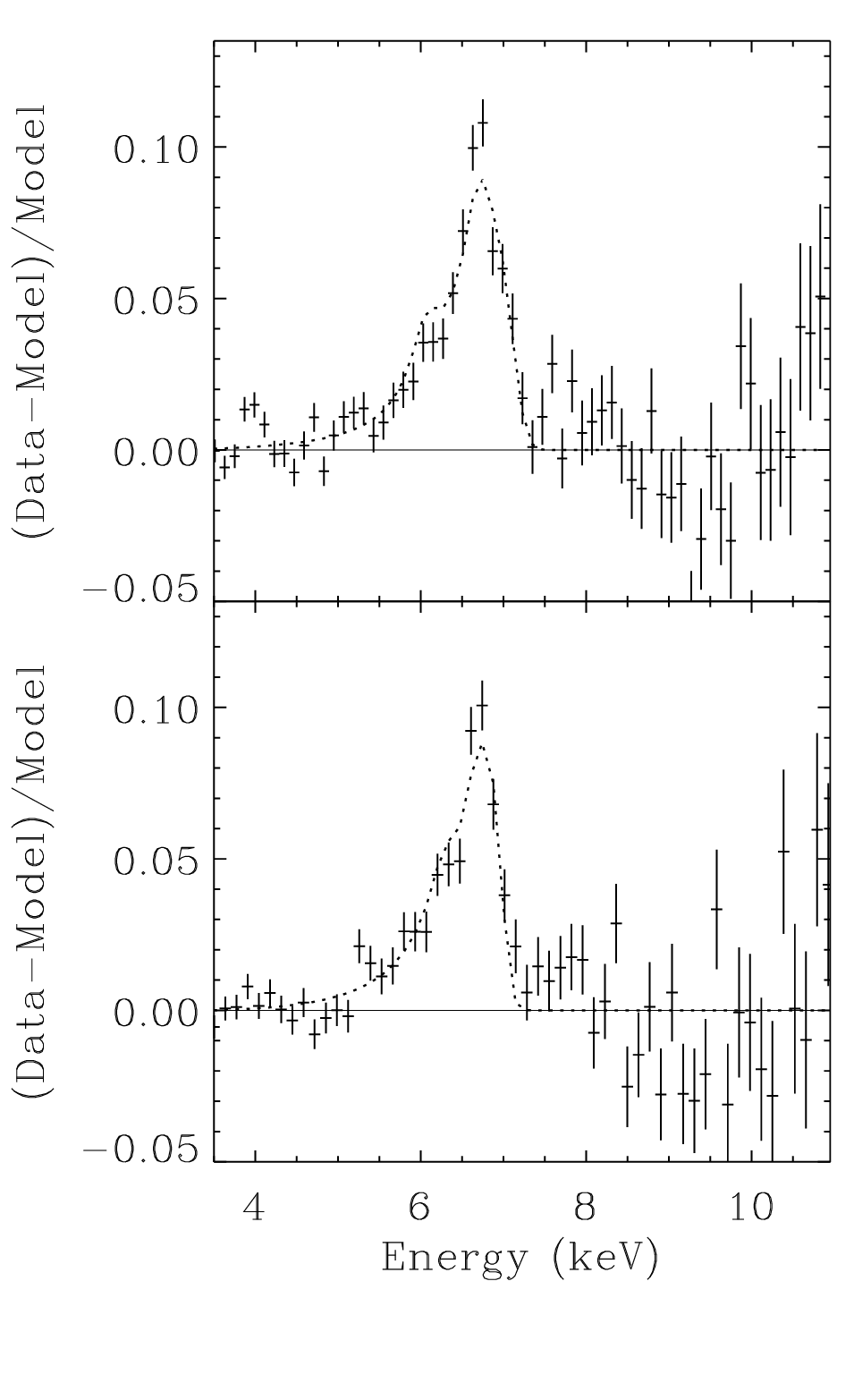}
\vspace{-0 cm}
\caption {{\it Upper panel}: {\it XMM-Newton} EPIC PN energy spectrum
from Ser X-1 for obs. 1 (obsId 0084020401).  With the error bars we
show the data intensity in excess of the model intensity, and
normalized by the model intensity (i.e., (Data$-$Model)/Model). Here
the model is the best-fit model (of Table 2) minus the {\tt laor}
component. The {\tt laor} component is separately shown with the
dotted line.  {\it Lower panel}: Similar to the upper panel, but for
obs. 3 (obsId 0084020601), and for the {\tt diskline} component (model
5 of Table 1) instead of the {\tt laor} component. For each panel, the
data points clearly show a broad relativistic iron emission line with
an extended red wing. }
\end{figure}

\end{document}